\documentstyle[twocolumn,prl,aps]{revtex}
\tightenlines
\begin{document}

\twocolumn[
\hsize\textwidth\columnwidth\hsize\csname@twocolumnfalse\endcsname
\draft

\title{Resonant Two-Magnon Raman Scattering in Cuprate
Antiferromagnetic Insulators
}
\author{
G.~Blumberg,$^{1,2,4}$ P.~Abbamonte,$^{2}$ M.~V.~Klein,$^{1,2}$
L.~L.~Miller,$^{3}$
W.~C.~Lee,$^{1,2,\dag}$ and D.~M.~Ginsberg$^{1,2}$
}
\address{
$^{1}$NSF Science and Technology Center for Superconductivity\\
$^{2}$Department of Physics,
University of Illinois at Urbana-Champaign, Urbana, IL 61801-3080\\
$^{3}$Ames Laboratory, Iowa State University, Ames IA 50011\\
$^{4}$Institute of Chemical Physics and Biophysics, R\"avala 10,
Tallinn EE0001, Estonia
}
\date{November 9, 1995; STCS-1153}
\maketitle

\begin{abstract}
We present results of low-temperature two-magnon resonance
Raman excitation profile measurements for single layer
Sr$_2$CuO$_2$Cl$_2$ and bilayer YBa$_2$Cu$_3$O$_{6 + \delta}$
antiferromagnets over the excitation region from 1.65
to 3.05~eV.  These data reveal composite structure of the
two-magnon line shape and strong nonmonotic dependence of the
scattering intensity on excitation energy.
We analyze these data using the {\em triple resonance} theory of Chubukov and
Frenkel (Phys. Rev. Lett., {\bf 74}, 3057
(1995)) and deduce information about magnetic
interaction and band parameters in these materials.

\end{abstract}

]
\narrowtext

The insulating phases of the high-$T_c$ materials are antiferromagnets
(AF) with spin ${\bf S} = \slantfrac{1}{2}$, localized on Cu atoms of
the CuO$_2$-planes, characterized by superexchange constant $J \approx
120$~meV and
a N\'{e}el temperature $T_N \approx 300$~K \cite{Vaknin87,Shirane87}.
The magnetic interactions in the CuO$_2$-planes can be described by a
Heisenberg Hamiltonian on the 2D square lattice
$H = J \sum_{<i,j>} ({\bf S}_i \cdot {\bf S}_j - \slantfrac{1}{4})$,
where ${\bf S}_i$ is the spin on site $i$ and
the summation is over nearest-neighbor Cu pairs.

Two-magnon (2M) Raman scattering (RS) probes mainly short wavelength magnetic
excitations in the AF lattice.
The scattering process involves a virtual charge-transfer excitation.
The traditional Hamiltonian for describing the interaction of light with spin
degrees of freedom is the Loudon-Fleury Hamiltonian \cite{Fleury68},
$H = \alpha \sum_{<i,j>} ({\bf \hat{e}}_i \cdot {\bf R}_{ij})
({\bf \hat{e}}_f \cdot {\bf R}_{ij}){\bf S}_i \cdot {\bf S}_j$,
where ${\bf \hat{e}}_i$ and ${\bf \hat{e}}_f$ are the polarization
vectors of incoming and outgoing photons, $\alpha$ is the coupling
constant and ${\bf R}_{ij}$
is a vector connecting two nearest-neighbor sites $i$ and $j$.
Shastry and Shraiman \cite{Shastry} have derived this
Hamiltonian in the large-$U$ Hubbard model approach for non-resonant excitation
energies.
Including final state interactions in this 2M Raman
scattering theory results in a peak with Raman frequency shift $\omega$
near $2.7-2.8J$ in
$B_{1g}$ scattering geometry for a single layer 2D lattice of $D_{4h}$ symmetry
\cite{ElliottParkinson,Singh91,Canali92,ChubukovLong95}.
This allows one to determine the
exchange interaction constant $J$ directly from the peak energy observed in
the Raman spectrum.
In fact, the Raman data from the $B_{1g}$ 2M scattering peak yielded the first
estimate of the exchange interaction constant in La$_2$CuO$_4$
\cite{LyonsPRB88,Sugai88,Tokura90}, Gd$_2$CuO$_4$ \cite{Sulewski91},
RBa$_2$Cu$_3$O$_{6 + \delta}$ (R=Y \cite{Lyons88}, Eu \cite{Knoll90},
Pr and Nd \cite{Yoshida90}),
Sr$_2$CuO$_2$Cl$_2$, LaGdCuO$_4$,
Nd$_2$CuO$_4$, (Ca,Sr)CuO$_2$ \cite{Tokura90}. The 2M peak has also
been observed in the electron-doped materials \cite{Tomeno91,Sugai91}
and in underdoped YBa$_2$Cu$_3$O$_{6 + \delta}$ and
YBa$_2$Cu$_4$O$_8$ superconductors \cite{Blumberg94}.
The temperature dependence of the 2M scattering has been studied
in Ref. \onlinecite{Knoll90}.

The dependence of 2M scattering intensity on the incoming excitation
energy $\omega_{i}$  has been discussed in a number of earlier works
\cite{Blumberg94,Lyons89,Sugai89,Sulewski90,Yoshida92,Liu93,Liu93error,Mayer94,Rubhausen95}.
Comparison of optical conductivity data with these Raman
measurements shows that
the maximum strength of the 2M scattering occurs for the
excitation energies substantially higher than the optical band
edge that marks the onset of particle-hole excitation
across the charge transfer gap (CTG).

In this Letter we report low-temperature two-magnon
resonance Raman excitation profile (RREP) measurements for single CuO$_{2}$
layer
Sr$_2$CuO$_2$Cl$_2$ and bilayer YBa$_2$Cu$_3$O$_{6 + \delta}$
AF insulators, covering the broad excitation region from 1.65 to 3.05~eV.
These data reveal composite structure of the 2M line shape and strong
nonmonotonic dependence of the shape and 2M scattering intensity on
the excitation energy $\omega_{i}$.
We analyze these results within the framework of
the Chubukov-Frenkel approach \cite{ChubukovLong95,Chubukov95}, show that
{\em triple resonance} is the leading process for resonant magnetic Raman
scattering, and extract information about the charge transfer band and
magnetic interactions in these cuprates.

The investigations were performed on single crystals grown as described
in Refs. \onlinecite{LeeGinsberg,Miller90}.
Spectra reported here were excited with different $Ar^+$, $Kr^+$ and
dye laser lines in $x'y'$, $x'x'$, $xy$ and $xx$
pseudo-back scattering geometries from an [001] crystal face.
With assumed $D_{4h}$ symmetry, this gives Raman
spectra of $B_{1g} + A_{2g}$, $A_{1g} + B_{2g}$, $B_{2g} + A_{2g}$ and
$B_{1g} + A_{1g}$ symmetry respectively. The $A_{2g}$ component in the
2M excitation region is known to be quite weak \cite{Sulewski91}.
The single crystals were mounted in a continuous helium flow optical
cryostat (Oxford Instruments) and kept at 5~K.
Less than 15~mW of the incident laser power
was focused in a 50~$\mu$m diameter spot on the
$ab$-plane crystal surface.
Light collected in pseudobackscattering
geometry was focused with achromatic optics (1:3 magnification) into the
0.5~mm (underfilled) entrance slit of a triple grating spectrometer (SPEX
1877). 150~lines/mm gratings were used to cover a broad spectral range.
The dispersed light was detected by a liquid nitrogen cooled 512 x 512
pixel CCD camera (Photometrics).
Spectral corrections to the detector sensitivity and
the frequency dependence of the collection optics and the spectrometer
system were made using a calibrated uniform light source sphere (USS-600
from Labsphere). We present the spectra in relative units of Raman cross
section per unit solid angle per unit volume per unit frequency shift.
The scattered intensity ($I_{f}$) is corrected to
obtain the cross section via
$d^2\sigma / d\omega d\Omega \propto I_{f} ( \alpha_{i} + \alpha_{f} ) n^2 /
( T_{i} T_{f} ( I_{i} / \omega_{i} ) ).$
This expression corrects for transmission coefficients at the sample-helium
interface
for the incident ($T_{i}$) and scattered ($T_{f}$) photons, absorption
of the incident ($\alpha_{i}$) and reabsorption of the scattered
($\alpha_{f}$)
photons and also the scattered photon solid angle correction at
the sample-helium interface.
These optical parameters were acquired from
polarized reflectivity and ellipsometry measurements.
We have used the 80~K optical data
from Refs.\onlinecite{Humlicek88,Zibold} for YBa$_2$Cu$_3$O$_{6 + \delta}$
and Sr$_2$CuO$_2$Cl$_2$ crystals respectively.
The maximum effect of these corrections in the energy range 1.5 to
3~eV was $\sim 300 \%$.

The Raman spectra excited with photon energies between 1.9 and 3.1~eV
in $x^{\prime}x^{\prime}$ and $x^{\prime}y^{\prime}$ geometries and dependence
of the $B_{1g}$ 2M peak intensity on the excitation energy are shown
in Figs.\ \ref{Raman} - \ref{RREP}.

The features of primary interest are the following:

(a) Resonant magnetic RS gives the main contribution in the
$B_{1g}$ channel.
However, as first shown by Sulewski {\em et al.}\cite{Sulewski91}, it
contributes to the $A_{1g}$ and $B_{2g}$ scattering geometries as well.
For these geometries RS bands are broad and have less structure.
The $A_{1g}$ intensity is weaker than in the $B_{1g}$ channel, and the $B_{2g}$
intensity is weaker than $A_{1g}$.

(b) In the $B_{1g}$ channel the 2M band is composite and contains
two asymmetric lineshapes.
Sr$_2$CuO$_2$Cl$_2$ shows a band with peaks at 2970
and 3900~cm$^{-1}$, where the second peak is well resolved for
certain excitation energies.
For YBa$_2$Cu$_3$O$_{6 + \delta}$ the 2M band is peaked at
2720~cm$^{-1}$, and the second peak at 3660~cm$^{-1}$ is resolved
only for the lower excitation energies.
The 2M band becomes much broader and more symmetric when the
excitation energy approaches 3~eV.

(c) In the $A_{1g}$ channel for Sr$_2$CuO$_2$Cl$_2$ the 2M band
intensity at lower frequencies is a combination of a linear and a cubic
power law.
For YBa$_2$Cu$_3$O$_{6 + \delta}$ the 2M band intensity has flat
low-frequency tail and a threshold at about 1750~cm$^{-1}$.
The sharp peaks at the low energy tails are due to resonant
multi-phonon scattering that becomes
strongly enhanced for excitations closed to the CTG energy.

(d) The 2M scattering intensity strongly depends on the incoming photon energy.
For YBa$_2$Cu$_3$O$_{6 + \delta}$ the $B_{1g}$ peak intensity at 2720~cm$^{-1}$
is enhanced by three orders of magnitude when the excitation
approaches 3~eV, that is about 1.3~eV higher than the first peak in
$\epsilon_{2}$ (See Fig. \ref{RREP}), and a weaker resonance is
observed at about 2.1~eV, $\sim0.32$~eV higher than the peak in $\epsilon_{2}$.
For Sr$_2$CuO$_2$Cl$_2$ the RREP at 2970~cm$^{-1}$ shows resonance
at 2.4~eV, about 0.44~eV higher than the $\epsilon_{2}$ peak
at 1.95~eV, and a steep rise to the second strong resonance in the UV region,
outside the accessibility of our experimental technique.

These results can be understood in terms of the 2M resonant
RS theory proposed by Chubukov and Frenkel
\cite{ChubukovLong95,Chubukov95} that suggests a leading resonant RS
diagram that is different from those leading to the Loudon-Fleury theory.
The proposed process is the following:
the incoming photon $\omega_{i}$ creates a virtual
electron-hole state consisting of an excitation across the CTG,
then the fermions
emit or absorb two magnons (one by each fermion)  with
momenta {\bf q} and {\bf -q} before recombining by emitting an outgoing photon
with the energy $\omega_{f} = \omega_{i} - \omega$.
The internal frequency integral of the corresponding diagram has
three denominators (See Fig. 2(b) and Eq. (5) in Ref. \cite{Chubukov95})
that could vanish simultaneously for a certain region of $\omega_{i}$
and $\omega_{f}$.
If the conditions for simultaneous vanishing of all three denominators are
satisfied, this process gives a resonant RS enhancement known as triple
resonance \cite{Cardona82}.
The {\em triple resonance} enhancement conditions depend on the dispersion
of the fermionic bands and of the magnons.
They have been calculated in Ref. \cite{Chubukov95} for the one-band
Hubbard model within the spin density wave formalism.
That model description of the electronic states at half filling
predicts no {\em triple resonance} conditions for excitation
frequencies $\omega_{i}$ around the CTG energy $2\Delta$.
The enhancement for Raman shifts $\omega$ between $\sim 2.4$ and $4J$ occurs
twice: first when $\omega_{i}$ lies between $\sim 2.4$ and $5.6J$ above
$2\Delta$ and second, even stronger, when $\omega_{i}$ approaches the
upper edge of the
electron-hole excitation band (See Fig. 3 in Ref.\cite{Chubukov95}).

This {\em triple resonance} process can explain each of the observed
features (a)-(d):

(a) The {\em triple resonance} contributes to all three experimentally observed
scattering symmetries: $B_{1g}$, $A_{1g}$ and $B_{2g}$.

(b) The largest contribution from {\em triple resonance} scattering is in the
$B_{1g}$ channel,
where, due to the geometric form factor for magnon emission, mainly
short wavelength magnons from the
vicinity of the magnetic Brillouin zone boundary
participate in the process.
The 2M density of states from the vicinity of these points diverges
at the maximum 2M energy $4J$.
Accounting for the final state interaction leads to partial renormalization
of the 2M spectral weight
\cite{ElliottParkinson,Singh91,Canali92,ChubukovLong95}
and shifts the maximum of the 2M spectral weight down to $\sim 2.8J$.
As a result, the observed 2M line shape (See Fig.~\ref{Raman}) consists of
two peaks: one is close to
$4J$, and the other, due to magnon-magnon scattering, is at about $2.8J$
($J_{SCOC} \approx 130$~meV,
$J_{YBCO} \approx 110$~meV) \cite{Jperp}.
The relative intensity of these two peaks depends on
the {\em triple resonance} conditions.

(c) In the $A_{1g}$ channel the long wavelength magnons from the vicinity
of the magnetic Brillouin zone center contribute to 2M
scattering and determine the low-frequency slope
of the 2M lineshape\cite{ChubukovDirk95} for Sr$_2$CuO$_2$Cl$_2$.
The threshold at about 1750~cm$^{-1}$ in the 2M intensity observed
for YBa$_2$Cu$_3$O$_{6 + \delta}$ could be the result
of interlayer superexchange $J_{\perp}$ within the CuO$_{2}$-plane
bilayers: the interlayer coupling results in an optical magnon branch with
a gap of $2\sqrt{J J_{\perp}}$ \cite{Tranquada89} and an onset for 2M intensity
at $4\sqrt{J J_{\perp}}$ \cite{ChubukovDirk95}.
The latter gives an estimate for $J_{\perp} \approx J/4$.

(d) The {\em triple resonance} theory predicts two peaks for the RREP at
$\omega = 2.8J$:
the first one at $\omega_{i} = \omega_{i}^{res1} \approx 2\Delta + 2.9J$, and
the
second at $\omega_{i}^{res2}$,
almost at the upper edge of the electron-hole excitation band.
This prediction correlates with observed features (See Fig.~\ref{RREP}),
where the
first resonance is about $2.9 - 3.4J$ above $2\Delta$, and the RREP shows
a strong rise to the second resonance.
Moreover, the model calculation of the Raman vertex in the absence of damping
predicts that the 2M peak intensity increases by an inverse linear
law as the incoming photon energy approaches the upper edge of the
charge-transfer excitation band.
In the inset of the Fig. \ref{RREP} we display the satisfactory linear fit of
the inverse 2M
intensity data for YBa$_2$Cu$_3$O$_{6 + \delta}$ versus $\omega_{i}$
and extract the linear singularity excitation energy
$\omega_{i}^{res2} \approx 3.1$~eV.
The latter allows one to estimate the electron-hole excitation band width as
$\omega_{i}^{res2} - 2\Delta \approx 1.3$~eV and nearest-neighbor
hopping integral $t_{YBCO} \approx 320$~meV since
$\omega_{i}^{res2} \approx 2\sqrt{\Delta^{2} + 16t^{2}}$.

Despite the agreement with experimental results, the {\em triple resonance}
calculations \cite{ChubukovLong95,Chubukov95} use a simplified effective
one-band Hubbard model which takes into account only nearest-neighbor
hopping integrals $t$.
By breaking electron-hole symmetry, the Hubbard model with
next-nearest-neighbour
hopping $t^{\prime}$ \cite{ChubukovMusaelian95}
provides satisfactory agreement between the narrow ($\approx 0.3$~eV)
valence band width observed in photoemission
experiments \cite{Wells95} and our estimate ($\approx 1.3$~eV) for the
sum of the valence and conduction bands width.
Accounting for $t^{\prime}$ also results
in a different strength and position of the {\em triple resonance}
at $\omega_{i}^{res1}$ in the two investigated AF's, but does not
effect  $\omega_{i}^{res2}$ and our estimate of the hopping integral
$t$ \cite{Chubukov3J}.

In summary, we have performed low-temperature two-magnon resonance
Raman excitation profile measurements for single layer
Sr$_2$CuO$_2$Cl$_2$ and bilayer YBa$_2$Cu$_3$O$_{6 + \delta}$ AF
insulators for the broad excitation region from 1.65 to 3.05~eV.
We have shown that the experimental data are in general agreement with the
theory proposed by Chubukov and Frenkel \cite{ChubukovLong95,Chubukov95},
indicating that the {\em triple resonance} process gives the leading
contribution to resonant 2M scattering.
This process contributes to $B_{1g}$, $A_{1g}$ and $B_{2g}$
scattering geometries.
The $B_{1g}$ 2M line shape is composite and contains two peaks at
about 2.8 and $4J$. The intensities of the two peaks strongly depend on the
excitation resonance conditions.
By measuring the excitation energy of the {\em triple resonance}
$\omega_{i}^{res2}$
and knowing the CTG from the optical conductivity,
we estimate the electron-hole excitation band width to be $\approx
1.3$~eV and the nearest-neighbor hopping integral
$t_{YBCO} \approx 320$~meV.
The $A_{1g}$ 2M band of bilayer YBa$_2$Cu$_3$O$_{6 + \delta}$ material
exhibits an optical gap that gives an estimate of interlayer
superexchange $J_{\perp} \approx J/4$.

We are indebted to A.\ V.\ Chubukov and D.\ M.\ Frenkel for many
valuable discussions.
We are grateful to J.\ Graybeal, D.\ Tanner, A.\ Zibold for providing
optical data.
This work was supported by NSF grant DMR 93-20892 (P.A.,
M.V.K.), cooperative agreement DMR 91-20000 through the STCS (G.B.,
M.V.K., W.C.L., D.M.G.),
and DOE W-7405-Eng82 through Ames Laboratory, ISU (L.L.M.).

\begin{figure}
\caption{
The two-magnon resonance Raman scattering cross sections
($x^{\prime}y^{\prime}$ and $x^{\prime}x^{\prime}$
geometries) for Sr$_2$CuO$_2$Cl$_2$ and YBa$_2$Cu$_3$O$_{6 + \delta}$
single crystals for different excitation energies above the charge
transfer gap $2\Delta$.
The scattering intensities for Sr$_2$CuO$_2$Cl$_2$ are
reduced by factor of 10 relative to YBa$_2$Cu$_3$O$_{6 + \delta}$.
Baselines are shifted as indicated.
Dashed lines mark $2.8J + J_{\perp}$ and $4J + J_{\perp}$ Raman shifts.
$J_{SCOC} = 1060$~cm$^{-1}$,
$J_{YBCO} = 895$~cm$^{-1}$,
$J_{\perp SCOC} = 0$,
$J_{\perp YBCO} = 215$~cm$^{-1}$.
}
\label{Raman}
\end{figure}

\begin{figure}
\caption{
The intensity of the $B_{1g}$ two-magnon peak at $2.8J$ (circles) as a function
of the excitation energy $\omega_{i}$.
Also shown is the imaginary part of the dielectric constant from
Refs.\protect\cite{Humlicek88,Zibold} (solid line).
Dashed lines mark the charge transfer gap energy $2\Delta$ and $2\Delta + 3J$.
Arrows indicate the {\em triple resonance} excitation energies.
The inset shows the linear fit (dashed line) to the inverse two-magnon peak
intensity (open circles). }
\label{RREP}
\end{figure}


\begin{references}
\bibitem{Liu93error}
The RREP data reported in Ref.
\protect\cite{Liu93}
for large $\omega_{i}$ is in error because of calibration difficulties.
\end{references}
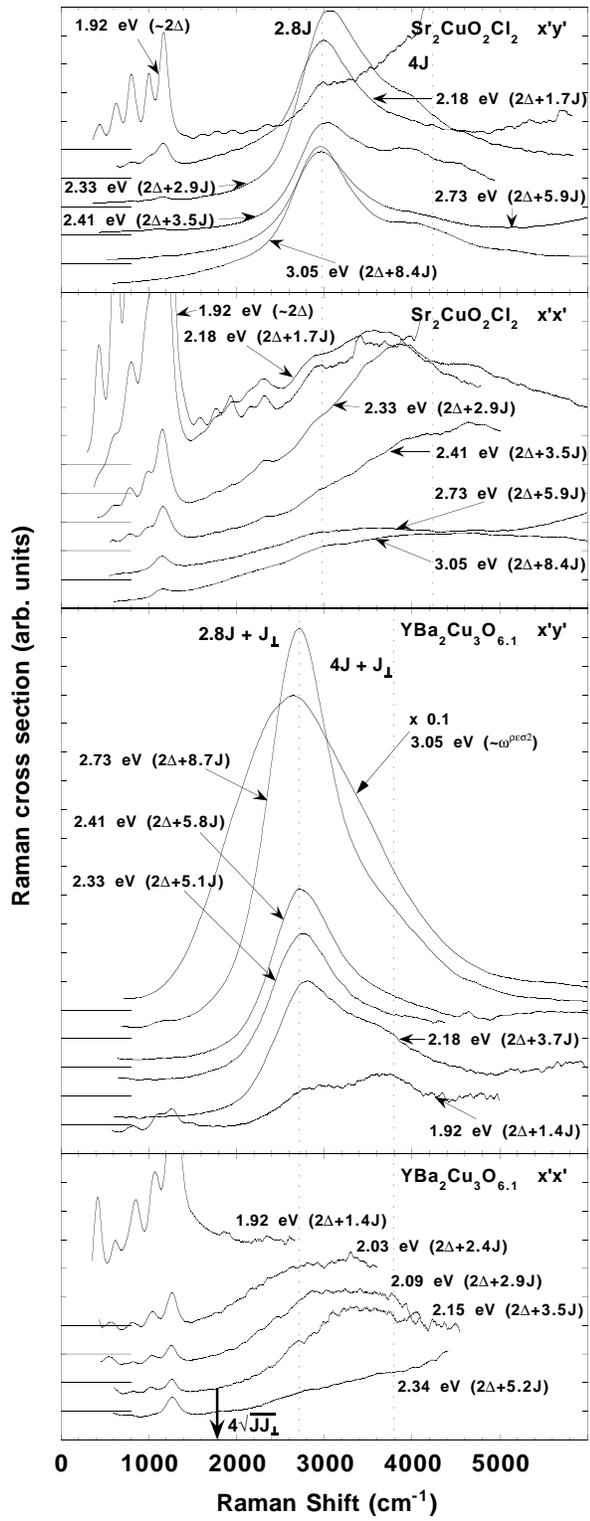
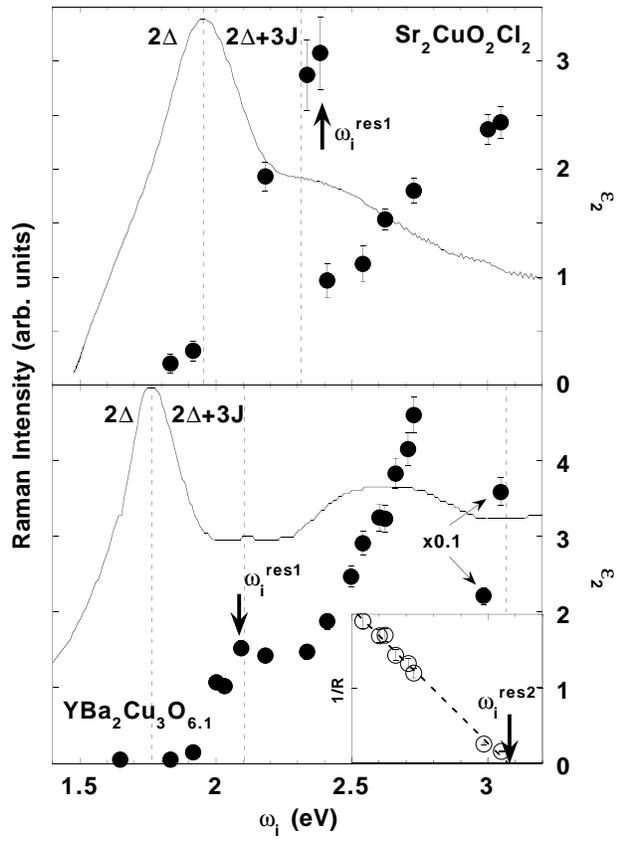
\end{document}